\documentclass[prb,twocolumn,aps,amsmath,amssymb,superscriptaddress]{revtex4-2}
\usepackage[T1]{fontenc}
\usepackage[utf8]{inputenc}
\usepackage{fancyhdr}
\usepackage{graphicx}
\usepackage{amsmath}
\usepackage{amssymb}
\usepackage{dcolumn}
\usepackage{float}
\usepackage{mathrsfs}
\usepackage{bm}
\usepackage{amsfonts,times}
\usepackage{adjustbox}
\usepackage{soul}
\usepackage{parskip}
\usepackage[breaklinks=true,colorlinks,citecolor=blue,linkcolor=blue,urlcolor=blue]{hyperref}

\DeclareMathAlphabet{\bi}{OML}{cmm}{b}{it}
\def\be{\begin{equation}}
\def\ee{\end{equation}}
\def\bearr{\begin{eqnarray}}
\def\eearr{\end{eqnarray}}

\begin{document}

\title{Parity splitting of coupled Kitaev sectors in a spinful $s$-wave quantum dot array}

\author{Mijanur Islam} 
\email{drmijanurphysics25@gmail.com}
\author{Mahan Mohseni}
\author{Daniel Miravet}
\author{Pawel Hawrylak}
\affiliation
{Department of Physics, University of Ottawa, Ottawa, Ontario, K1N 6N5, Canada.}

\begin{abstract}
We discuss the  Majorana Fermions and parity splitting in spinful semiconductor-$s$-type-superconductor quantum dot arrays. Quantum dot arrays are described by interdot tunneling, spin orbit coupling, Zeeman splitting, and $s$-type pairing. We reorganize this Hamiltonian into two effective coupled Hamiltonians of pseudospin up and down Kitaev like chains. Using a bond fermion representation, we analyze the energy spectrum, parity resolved states, pseudospin polarization, and local electron addition spectral function. We determine  parameters for which one effective pseudospin  sector satisfies the Kitaev condition and supports Majorana zero modes.  We describe the coupling between the two pseudospin sectors that hybridizes Majorana modes, lifts the parity degeneracy, and shifts their spectral weight to finite energy. 
\end{abstract}

\maketitle

\section{Introduction}
Topological superconductors (TSs) hosting Majorana zero modes (MZMs) offer a promising platform for fault tolerant quantum computation through nonlocal encoding of information and non Abelian exchange statistics \cite{Kitaev2001,Nayak2008,Alicea2012,Beenakker2013,Aasen2016, Prada2020, Tsintzis2024}. One of the physical  realizations of Majorana zero modes is the Kitaev chain \cite{Kitaev2001}. The Kitaev chain describes an array of quantum dots (QDs) with spinless fermions, with nearest neighbor tunneling and $p$-wave  superconductive pairing. In topological phase Kitaev chain supports pairs of Majorana modes with zero energy localized at the edges of the chain \cite{Kitaev2001}. Further, braiding of MZM is insensitive to the details of external potential and leads to MZM with long coherence times \cite{Nayak2008,Alicea2011,Aasen2016}. Semiconductor-superconductor (SM-SC) arrays  with proximity induced $s$-wave pairing contain spinful electrons \cite{BCS1957,Tinkham,DeGennes}. An effective spinless $p$-wave description can emerge through the combined effects of Rashba spin orbit coupling (SOC), Zeeman splitting, and superconductivity \cite{Lutchyn2010,Oreg2010,Alicea2012,Prada2020}. Establishing this connection at the microscopic level is important because zero bias spectroscopic features alone cannot distinguish topological MZMs from conventional Andreev bound states (ABSs) arising from disorder or smooth confinement \cite{Pan2020,DasSarma2021,Prada2020}.

Semiconductor quantum dot arrays can be realized either by growth of QDs in semiconductor nanowires \cite{Ciorga2000, Dalacu2012,Phoenix2022} or by gating \cite{CIORGA2001, Ciorga2002, Altintas2021, gaudreau2012}. Semiconductor gated QD arrays provide a tunable route to artificial Kitaev chains, with electrostatic control over dot energies and effective interdot couplings \cite{Sau2012,Leijnse2012,Su2017,Ptok2018}. In dots coupled through superconducting regions, elastic cotunneling and crossed Andreev reflection generate effective hopping and nonlocal pairing \cite{Leijnse2012, Samuelson2024}, respectively. Balancing these processes can produce a sweet spot with nearly degenerate lowest even and odd parity states and spatially separated low energy modes. Experiments on two coupled QDs in InSb nanowire have demonstrated such states, commonly called ``poor man's Majoranas'' \cite{Dvir2022,Haaf2024}. The localization, robustness, excitation gap, and topological connection of these modes \cite{Tsintzis2022Creating,Zatelli2024,Liu2024Gap,Luethi2025,Dourado2025} have been explored. The three site chains show improved stability and spatial separation \cite{Bordin2025,Bordin2026,Haaf2025}. Advances in parity readout, phase control, and automated sweet spot tuning \cite{Benestad2024,Yang2025,VanLoo2025,krawczyk2026} further motivate microscopic descriptions that retain the spinful structure of the QDs. These developments allow to retain the underlying spinful degrees of freedom and clarify their relation to the ideal Kitaev model.

InAsP QDs embedded in wurtzite InP nanowires provide an alternative platform for semiconductor QD arrays, with demonstrated growth control over dot geometry, composition, and axial separation \cite{Dalacu2012,Phoenix2022}. Their electronic structure and interdot tunnel couplings have been investigated experimentally and theoretically \cite{Dalacu2012,Cygorek2020,Laferriere2021,Manalo2021,Phoenix2022,Manalo2024}. Previous work has explored Kitaev chain physics in InAsP/InP arrays coupled to a $p$-wave superconductor \cite{Mohseni2023,Mohseni2025}. A lack of easily attainable $p$-wave SC motivates examining the effective chain description when the  proximity induced pairing is instead of the $s$-wave type.

Here, we consider a QD chain with electrons with spin up and spin down with nearest neighbor QD-QD hopping, Rashba SOC, and Zeeman splitting, proximity induced on site $s$-wave pairing \cite{Stoudenmire2011,Pan2023}. We show that a fermionic transformation reorganizes the Hamiltonian into two coupled Kitaev like pseudospin sectors. We examine how the coupling hybridizes the nominal Majorana modes of the individual sectors and produces finite parity splitting in the spinful system. Energy spectra, parity resolved states, pseudospin sector polarization, and local electron addition spectral functions characterize the resulting low energy excitations and their spatial distribution.

The paper is organized as follows. Section~\ref{Hamiltonian} introduces the model. Section~\ref{Formalism} describes the fermionic transformations, Majorana, and bond fermion representations. Section~\ref{Results} presents the results, and Section~\ref{Conclusions} summarizes our findings and discusses future work.

\section{Model Hamiltonian}
\label{Hamiltonian}
\begin{figure}
    \centering
    \includegraphics[width=\linewidth]{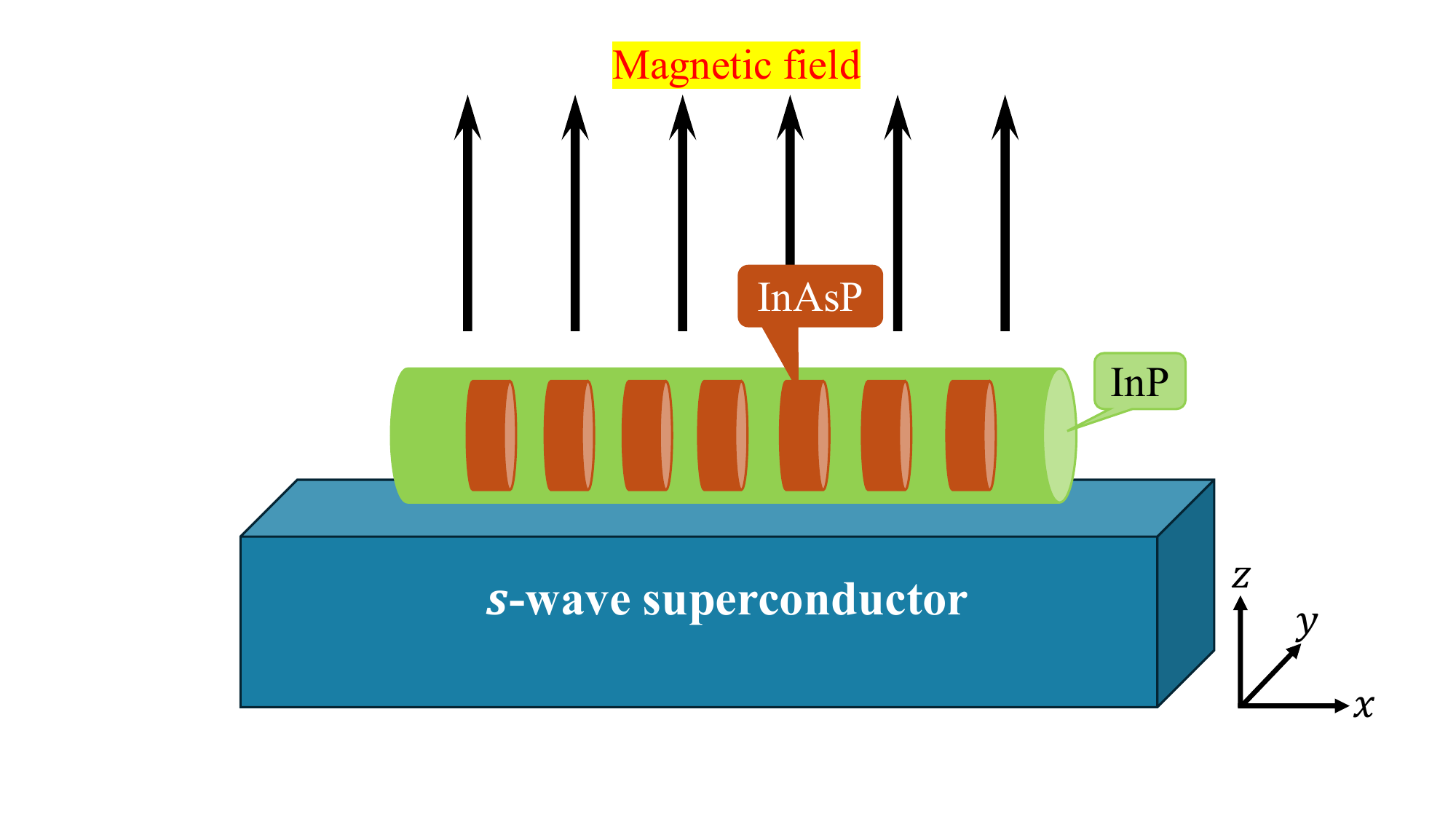}
    \caption{Schematic illustration of an InP nanowire hosting an array of  InAsP quantum dots, proximity coupled to an $s$-wave superconductor and subjected to a perpendicular magnetic field, chosen to be orthogonal to the effective Rashba spin orbit coupling and thereby inducing Zeeman splitting.}
    \label{fig:system}
\end{figure}

As discussed earlier, we consider a hexagonal InP nanowire containing an array of embedded InAsP quantum dots~\cite{Cygorek2020,Manalo2021,Dalacu2012,Phoenix2022,northeast2021,jaworowski2017,Koong2020} as shown in Fig. \ref{fig:system}. Electrons are confined to the QDs in nanowire. Electrons tunnel from dot to dot conserving spin direction and flipping spin due to Rashba SOC of strength $\alpha$.  A Zeeman splitting $V_z$ is produced by an externally applied magnetic field. The field is taken orthogonal to the effective spin orbit field (e.g. perpendicular to the spin quantization axis induced by Rashba SOC) so that it opens an energy gap. Due to proximity to an $s$-wave superconductor spin up and spin down Cooper pairs tunnel into QDs inducing a pairing amplitude $\Delta$ in each QD~\cite{Stoudenmire2011}. The quantum dot array is modeled as a tight binding chain of localized quantum dot orbitals coupled by tunnel amplitudes. Such arrays have been widely investigated~\cite{Cygorek2020,Manalo2021,Dalacu2012,Phoenix2022}. In the tight binding approximation, the Hamiltonian for this system can be written as~\cite{Stoudenmire2011,Souto2023, Tsintzis2024}
\begin{equation}
\label{eq:KC}
    H_e = H_{\textrm{TB}} + H_\mu + H_{\textrm{SOC}} + H_z + H_\Delta + H_U + H_V,
\end{equation}
where
\begin{align}
    \label{eq:tb}
        H_{\textrm{TB}} &= t\sum_{j=1, \sigma=\uparrow,\downarrow}^{N-1}\big(c_{j+1 \sigma}^\dagger c_{j\sigma} + h.c.\big),
    \\
    \label{eq:mu}
        H_\mu &= - \mu \sum_{j=1,\sigma=\uparrow,\downarrow}^N n_{j\sigma},
    \\
    \label{eq:soc}
        H_{\textrm{SOC}} &= \alpha \sum_{j=1}^{N-1}\big(c_{j+1 \uparrow}^\dagger c_{j\downarrow} - c_{j+1\downarrow}^\dagger c_{j\uparrow} +h.c. \big),
    \\
    \label{eq:vz}
        H_{z} &= V_z \sum_{j=1}^N \big( c_{j\uparrow}^\dagger c_{j\uparrow}- c_{j\downarrow}^\dagger c_{j\downarrow} \big),
    \\
    \label{eq:delta}
        H_\Delta &= \Delta \sum_{j=1}^N \big( c_{j\downarrow}^\dagger c_{j\uparrow}^\dagger + h.c. \big),
    \\
    \label{eq:Hubbard}
        H_U &= U \sum_{j=1}^N n_{j\uparrow}n_{j\downarrow},
    \\
    \label{eq:exthubbard}
        H_V &= V \sum_{j=1}^{N-1} n_j n_{j+1}. 
\end{align}
Here $c_{j\sigma}^\dagger$ creates an electron in the dot $j$ with spin $\sigma$, $n_{j\sigma}=c_{j\sigma}^\dagger c_{j\sigma}$ denotes the corresponding local spin density operator, and $n_j = \sum_{\sigma = \uparrow,\downarrow} n_{j\sigma}$ is the charge density, $t$ is the nearest neighbor hopping amplitude, $\mu$ is the onsite chemical potential. The SOC term is written in a gauge where the spin dependent hopping is proportional to $\sigma_y$. The magnetic field is accounted by Zeeman interaction and chosen along $\sigma_z$. Further, $U$ is the onsite short range Hubbard interaction and V is the nearest neighbor interdot density-density repulsive interaction. In the following, we focus on a description of the nanowire quantum dot array with SC pairing as the most important electron-electron interaction effect. Although Coulomb interactions~\cite{Stoudenmire2011} and charging effects can be significant in isolated InAsP/InP quantum dots~\cite{Cygorek2020,Manalo2021,Manalo2024,Allami2025}, they are neglected here to isolate the physics captured by the non interacting Hamiltonian. The effect of electron-electron Coulomb interaction will be reported in~\cite{Islam2026prep}.

\section{Formalism}
\label{Formalism}
\subsection{Mapping onto two coupled Kitaev like sectors}
\label{sec:onsite_diagonalization}

We will now  isolate  the on-site Hamiltonian to introduce a new set of quasi-particles mixing spin up and down fermions into pseudospin. The new quasi-particles provide a transparent basis in which the on-site Hamiltonian is diagonal, and the remaining hopping and SOC terms acquire renormalized amplitudes, pseudospin, and induced nonlocal pairing terms.

We begin by isolating the on-site terms of the electronic Hamiltonian (Eq. (\ref{eq:KC})), containing the Zeeman field (Eq. (\ref{eq:vz})), chemical potential (Eq. (\ref{eq:mu})), and $s$-wave pairing (Eq. (\ref{eq:delta})):
\begin{align}
\label{eq:H_onsite_original}
H_{\mathrm{onsite}}
=&
\sum_{j=1}^{N}
\Big[
(V_z-\mu)c_{j\uparrow}^\dagger c_{j\uparrow}
-(V_z+\mu)c_{j\downarrow}^\dagger c_{j\downarrow} \nonumber \\
&+ \Delta \, ( c_{j\downarrow}^\dagger c_{j\uparrow}^\dagger + c_{j\uparrow} c_{j\downarrow} )
\Big].%
\end{align}
It is convenient to rewrite Eq.~(\ref{eq:H_onsite_original}) in the Nambu spinor basis
$\big(c_{j\uparrow},\, c_{j\downarrow}^\dagger\big)^T$ as
\begin{equation}
H_{\mathrm{onsite}} = \sum_{j=1}^{N}
\begin{pmatrix}
c_{j\uparrow}^\dagger & c_{j\downarrow}
\end{pmatrix}
[A]
\begin{pmatrix}
c_{j\uparrow} \\
c_{j\downarrow}^\dagger
\end{pmatrix}
- N(V_z+\mu),
\label{eq:H_onsite_original1}
\end{equation}
where the on-site Bogoliubov--de Gennes (BdG) matrix is
\begin{equation}
A = V_z I - \mu \sigma_z - \Delta \sigma_x.
\label{eq:A_matrix}
\end{equation}
The eigenvalues of Eq.~\eqref{eq:A_matrix} are $\lambda_{\pm} = V_z \pm \tilde{\mu}$ where $\tilde{\mu}= \sqrt{\mu^2+\Delta^2}$. Diagonalizing Eq.~\eqref{eq:A_matrix} defines new  fermions $d_{j\sigma}$
\begin{equation}
\label{eq:trans}
\begin{pmatrix}
c_{j\uparrow} \\
c_{j\downarrow}^\dagger
\end{pmatrix}
= \chi_+ d_{j\uparrow} + \chi_- d_{j\downarrow}^\dagger
= \beta
\begin{pmatrix}
-\xi d_{j\uparrow} + d_{j\downarrow}^\dagger \\
d_{j\uparrow} + \xi d_{j\downarrow}^\dagger
\end{pmatrix},
\end{equation}
with $\chi_+ = \beta \begin{pmatrix} -\xi \\ 1 \end{pmatrix}$ and $\chi_- = \beta\begin{pmatrix} 1 \\ \xi \end{pmatrix}$, where $\xi = \frac{\Delta}{\mu+\sqrt{\mu^2+\Delta^2}}$ and $\beta = \frac{1}{\sqrt{1+\xi^2}}$. The $d$-fermions satisfy the canonical fermionic anticommutation relations $\{d_{i\sigma},d_{j\sigma'}^\dagger\} = \delta_{ij}\delta_{\sigma\sigma'}$. Substituting the transformation (Eq. (\ref{eq:trans})) into Eq.~(\ref{eq:H_onsite_original1}) gives
\begin{equation}
\resizebox{0.49\textwidth}{!}{
$H_{\mathrm{onsite}}
= \sum_{j=1}^N \left[\lambda_+ d_{j\uparrow}^\dagger d_{j\uparrow} -\lambda_- d_{j\downarrow}^\dagger d_{j\downarrow} \right] - N\left(\mu + \tilde{\mu}\right)$,}
\end{equation}
which is diagonal in the $d$-fermion basis.

\begin{figure*}[t]
    \centering
    \includegraphics[width=0.95\textwidth]{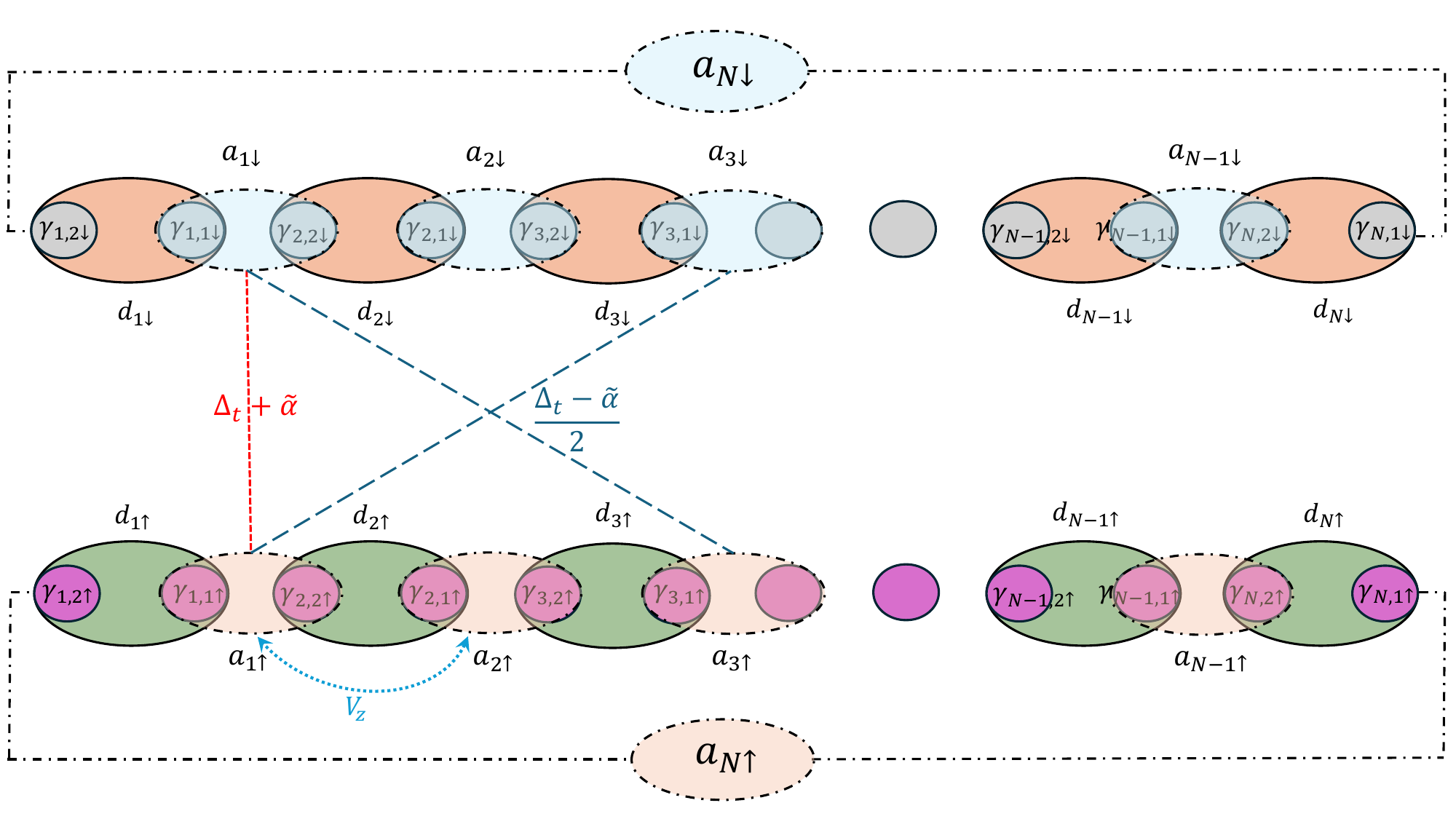}
    \caption{Schematic of the two coupled Kitaev chains (upper one is for pseudo down spin and lower one is for pseudo up spin) in the Majorana and bond fermion representation, with non-zero bond fermions $a_{j\downarrow}$ and $a_{j\uparrow}$ $(j\neq N)$, and the non local zero mode, living on the two ends of the two chains represent as $a_{N\downarrow}$ and $a_{N\uparrow}$. The coupling between the two chains are shown in red and blue dashed lines.}
    \label{fig:schematic}
\end{figure*}

The remaining terms of the Hamiltonian (\ref{eq:KC}), nearest neighbor hopping (Eq. (\ref{eq:tb})) and Rashba SOC term (Eq. (\ref{eq:soc})) can be rewritten in terms of the $d$ fermions. The hopping term (Eq. (\ref{eq:tb})) transforms as
\begin{align}
    H_{\mathrm{TB}}
=& \, \tilde{t}\sum_{j=1}^{N-1}[d_{j+1\uparrow}^\dagger d_{j\uparrow} + d_{j+1\downarrow}^\dagger d_{j\downarrow} + h.c.] \nonumber \\
& + \Delta_t\sum_{j=1}^{N-1}[d_{j+1\uparrow}^\dagger d_{j\downarrow}^\dagger - d_{j+1\downarrow}^\dagger d_{j\uparrow}^\dagger + h.c.]
\end{align}
with new effective parameters $\tilde{t} = -\frac{t\mu}{\tilde{\mu}},\; \Delta_t = -\frac{t\Delta}{\tilde{\mu}}$. Thus, the hopping term results in a renormalized hopping plus a superconducting like pairing between $d$-fermions on different lattice sites. Importantly, the Rashba SOC term (Eq. (\ref{eq:soc}))  transforms to generate a $p$-wave pairing as
\begin{align}
    H_{\mathrm{SOC}}
=& \, \tilde{\alpha}\sum_{j=1}^{N-1}[d_{j+1\uparrow}^\dagger d_{j\downarrow} - d_{j+1\downarrow}^\dagger d_{j\uparrow} + h.c.] \nonumber \\
& + \Delta_\alpha\sum_{j=1}^{N-1}[d_{j+1\uparrow}^\dagger d_{j\uparrow}^\dagger + d_{j+1\downarrow}^\dagger d_{j\downarrow}^\dagger + h.c.]
\end{align}
with $\tilde{\alpha} = \frac{\alpha\mu}{\tilde{\mu}},\; \Delta_\alpha = -\frac{\alpha \Delta}{\tilde{\mu}}$.
Therefore, here  we ended up with  a $p$-wave like pairing term and with a renormalized spin-orbit coupling. 

Collecting all contributions in the new operator basis, the full electronic Hamiltonian, Eq. (\ref{eq:KC}), can be written as 
\begin{equation}
\label{eq:H_total}
    H_e = H_{\downarrow} + H_{\uparrow }+ H_{\downarrow \uparrow}
- N\left(\mu + \tilde{\mu}\right),
\end{equation}
where the three components are
\begin{subequations}
\label{eq:effective}
\begin{align}
\label{eq:dnspin}
    H_{\sigma } =& \, \tilde{t}\sum_{j=1}^{N-1} [d_{j+1\sigma}^\dagger d_{j\sigma} + h.c.] + \Delta_\alpha \sum_{j=1}^{N-1}[d_{j+1 \sigma}^\dagger d_{j\sigma}^\dagger + h.c.] \nonumber \\
    & \pm \lambda_{\pm} \sum_{j=1}^N d_{j\sigma}^\dagger d_{j\sigma},
\end{align}
\begin{align}
    \label{eq:mixed}
    H_{\downarrow \uparrow} = & \, \Delta_t \sum_{j=1}^{N-1} [d_{j+1\uparrow}^\dagger d_{j\downarrow}^\dagger - d_{j+1\downarrow}^\dagger d_{j\uparrow}^\dagger + h.c.] \nonumber \\
    & + \tilde{\alpha} \sum_{j=1}^{N-1}[d_{j+1 \uparrow}^\dagger d_{j\downarrow} - d_{j+1 \downarrow}^\dagger d_{j\uparrow} + h.c.].
\end{align}
\end{subequations}

Here, $\sigma$ labels the pseudospin degree of freedom, with $+$ and $-$ denoting the pseudo up-spin and down-spin sectors, respectively. The transformed Hamiltonian separates into two Kitaev sectors labeled by pseudospins and  an inter pseudospin coupling $H_{\downarrow \uparrow}$, containing both induced nonlocal pairing and Kitaev sector mixing terms.

\subsection{Majorana fermion representation}
 We now express the Hamiltonians in Eq. (\ref{eq:effective}) in terms of Majorana and bond fermions. The schematic representation is shown in Fig. \ref{fig:schematic}. To this end, each fermionic operator $d_{j\sigma}$  and its Hermitian conjugate  $d_{j\sigma}^\dagger$ is decomposed into two Majorana fermion operators, $\gamma_{j,1\sigma}$ and $\gamma_{j,2\sigma}$~\cite{Kitaev2001, Mohseni2023}, according to
\begin{equation}
\label{eq:majo}
    \gamma_{j,1\sigma} = (d_{j\sigma} + d_{j\sigma}^\dagger), \;\;\;\;\;\;\; \gamma_{j,2\sigma} = i(d_{j\sigma} - d_{j\sigma}^\dagger).
\end{equation}
These operators are Hermitian, $ \gamma_{j,a\sigma}^\dagger = \gamma_{j,a\sigma} $ and satisfy the Majorana anti-commutation relations $\{\gamma_{i,a\sigma}, \gamma_{j,b\sigma^\prime} \} = 2\delta_{ij}\delta_{ab}\delta_{\sigma \sigma^\prime}$.

Using Eq.~(\ref{eq:majo}) together with the above anticommutation relations, the Hamiltonians in Eq.~(\ref{eq:effective}) can be rewritten in terms of Majorana fermionic operators as 
\begin{widetext}
\begin{subequations}
\label{eq:effective_majo}
\begin{equation}
\label{eq:dnspin_majo}
H_\sigma = \frac{i}{2}\bigg[(\tilde{t} + \Delta_\alpha)\sum_{j=1}^{N-1} \gamma_{j+1,2\sigma} \gamma_{j,1\sigma} + (\tilde{t} - \Delta_\alpha) \sum_{j=1}^{N-1}\gamma_{j,2\sigma} \gamma_{j+1,1\sigma} \pm \lambda_\pm \sum_{j=1}^N (\gamma_{j,2\sigma} \gamma_{j,1\sigma} - i)\bigg],
\end{equation}
\begin{equation}
\label{eq:mixed_majo}
H_{\downarrow \uparrow} = \frac{i(\Delta_t + \tilde{\alpha})}{2} \sum_{j=1}^{N-1} \bigg[\gamma_{j+1,2\uparrow} \gamma_{j,1\downarrow} - \gamma_{j+1,2\downarrow} \gamma_{j,1\uparrow}\bigg] + \frac{i(\Delta_t - \tilde{\alpha})}{2} \sum_{j=1}^{N-1}\bigg[\gamma_{j+1,1\uparrow} \gamma_{j,2\downarrow} - \gamma_{j+1,1\downarrow} \gamma_{j,2\uparrow}\bigg].
\end{equation}
\end{subequations}
\end{widetext}
This representation makes the structure of hopping and pairing processes explicit in the Majorana representation  and motivates the introduction of new fermions, which combine neighboring Majorana operators into conventional fermionic degrees of freedom. The Hamiltonian in Eq.~(\ref{eq:effective_majo}) explicitly exhibits couplings between different types of Majorana operators on adjacent sites, both within a given Kitaev sector and between opposite sectors.

\subsection{Bond fermion representation}
The specific pattern of these inter sector Majorana pairings suggests a reorganization of the degrees of freedom. This observation motivates the definition of a new set of pseudo fermionic operators, referred to as {\it bond fermions}~\cite{Mohseni2023}, constructed from pairs of Majorana operators of different types on neighboring sites, as defined below
\begin{equation}
a_{j\sigma} =
\begin{cases}
\dfrac{1}{2}\bigl(\gamma_{j,1\sigma} - i \gamma_{j+1,2\sigma}\bigr),
& j = 1,\dots,N-1, \\[6pt]
\dfrac{1}{2}\bigl(\gamma_{N,1\sigma} - i \gamma_{1,2\sigma}\bigr),
& j = N .
\end{cases}
\end{equation}
Here, \(a_{j\sigma}\) denotes a bond fermion constructed from two Majorana operators of different types on neighboring sites when $j\neq N$. The operator \(a_{N\sigma}\) is defined by pairing the two Majorana operators \(\gamma_{1,2\sigma}\) and \(\gamma_{N,1\sigma}\), which remain unpaired at the ends of the chain. This fermionic operator corresponds to the zero-energy degree of freedom formed from the Majorana end modes in the topological phase. We focus on the Kitaev sweet spot of the pseudo down spin sector, defined by $\tilde{t} = \Delta_\alpha$ and $\lambda_- = 0$, or equivalently $t\mu = \alpha \Delta$ and $V_z = \sqrt{\mu^2 + \Delta^2}$. For the QD array described by Eq.~(\ref{eq:KC}), the Rashba spin orbit coupling is proportional to $\sin k$, where $k$ is the momentum and therefore vanishes at $k=0$ and $k=\pi$. The bulk topological phase is characterized by a negative product of the Pfaffians at these two momenta,
\begin{align*}
   \big[(2t-\mu)^2+\Delta^2-V_z^2\big]
\big[(2t+\mu)^2+\Delta^2-V_z^2\big]<0. 
\end{align*}
Substituting the sweet spot value reduces this inequality to \(16t^2(t^2-\mu^2)<0\). Using $t\mu = \alpha\Delta$, we thus obtain $|\alpha\Delta|>t^2$, or $\alpha\Delta>t^2$ for the positive parameters used here. Under these conditions, $H_\downarrow$ becomes diagonal in the bond fermion basis. Consequently, the Hamiltonians in Eq.~(\ref{eq:effective_majo}) written in the bond fermion basis take the form
\begin{widetext}
\begin{subequations}
\label{eq:effective_bond}
\begin{equation}
\label{eq:dnspin}
H_\downarrow =\tilde{t} \sum_{j=1}^{N-1} \bigg(2a_{j\downarrow}^\dagger a_{j\downarrow}-1\bigg) + 0a_{N\downarrow}^\dagger a_{N\downarrow}
\end{equation}
\begin{equation}
\label{eq:upspin}
H_\uparrow = \tilde{t} \sum_{j=1}^{N-1} \bigg(2a_{j\uparrow}^\dagger a_{j\uparrow}-1\bigg) + 0a_{N\uparrow}^\dagger a_{N\uparrow} + V_z\sum_{j=1}^{N}\big(a_{j-1\uparrow}^\dagger a_{j\uparrow}^\dagger + a_{j-1\uparrow}^\dagger a_{j\uparrow} + h.c.\big) + NV_z
\end{equation}
\begin{equation}
\label{eq:mixed}
H_{\downarrow \uparrow} = (\Delta_t + \tilde{\alpha}) \sum_{j=1}^{N-1} \big(a_{j\uparrow}^\dagger a_{j\downarrow}^\dagger +h.c.\big) + \frac{(\Delta_t - \tilde{\alpha})}{2} \sum_{j=1}^{N-1}\big(a_{j+1\uparrow}^\dagger a_{j-1\downarrow}^\dagger - a_{j+1\downarrow}^\dagger a_{j-1\uparrow}^\dagger - a_{j+1\uparrow}^\dagger a_{j-1\downarrow} + a_{j+1\downarrow}^\dagger a_{j-1\uparrow} + h.c. \big),
\end{equation}
\end{subequations}
\end{widetext}
where in Eqs. (\ref{eq:upspin}) and (\ref{eq:mixed}) one should identify $a_{0\sigma} \equiv a_{N\sigma}$. The total Hamiltonian in terms of bond fermion can be expressed via Eq. (\ref{eq:H_total}). Furthermore, an $s$-wave Majorana nanowire can be interpreted as two effective $p$-wave Kitaev chains associated with different pseudospin species. The two pseudospin chains are coupled through both short and long range superconducting pairing and Rashba spin orbit coupling.

Following the procedure outlined in Ref.~\cite{Mohseni2023}, we employ the bond fermion occupation basis to perform exact diagonalization of the total Hamiltonian. The goal of this work is to develop a bond fermion representation that provides a transparent way to analyze the  transformed Hamiltonian and spatial structure of its eigenvectors. In this representation, the emergence of a decoupled or weakly coupled bond fermion at the sweet spot condition offers a useful Hamiltonian level diagnostic of Majorana  edge localised states. We support the bond fermion analysis with numerical calculations of the energy spectra, parity splitting, and spectral functions. Within this combined framework, the bond fermion formulation provides an intuitive description of how edge localised states (MZM) emerge, hybridize, and respond to coupling between the two Kitaev like sectors.

\section{Results and discussion}
\label{Results}
\begin{figure}
    \centering
    \includegraphics[width=\linewidth]{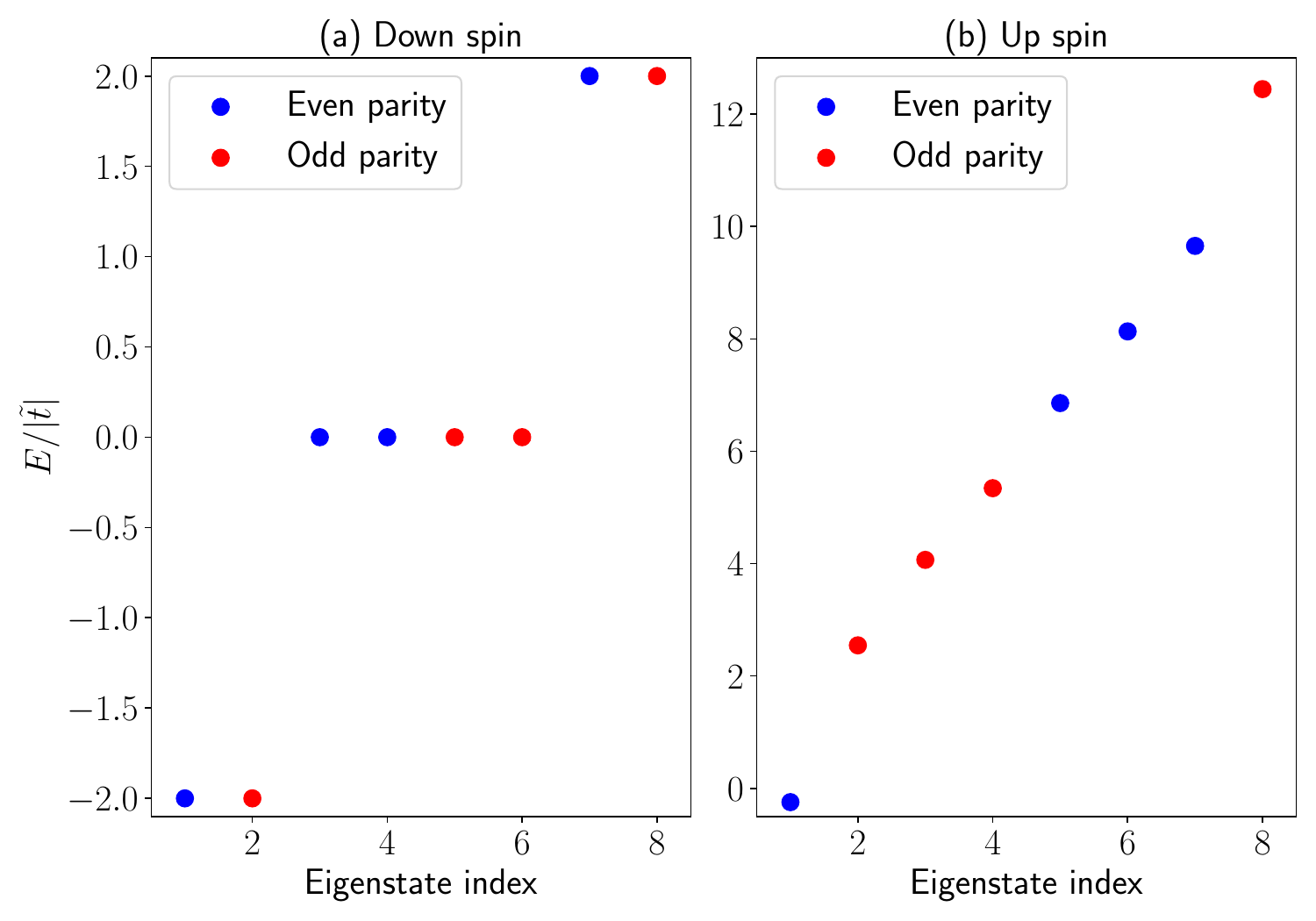}
    \caption{Energy spectra of the pseudo (a) down and (b) up spin chains in the bond fermion basis, where we set the pseudo down spin chain to be topological, imposing $\tilde{t} = \Delta_\alpha$ and $V_z = \sqrt{\mu^2 + \Delta^2}$. The parameters are $N=3$, $t=\Delta=1$ and $\alpha=1.2t$, consequently $\tilde{t}=-0.7682$ and $V_z = 1.5620$. Energies are normalized to $|\tilde{t}|$.}
    \label{fig:Energy_spin}
\end{figure}

We first examine the energy spectra of the pseudo down-spin ($H_\downarrow$) and up-spin ($H_\uparrow$) sectors separately. In the bond fermion representation, the topological phase is signaled by the presence of an unpaired zero energy bond fermion, localized at the boundaries. This corresponds to a bond fermion that remains decoupled from the Hamiltonian and therefore does not acquire a finite excitation energy. This structure follows from Eqs.~(\ref{eq:dnspin}) and (\ref{eq:upspin}), in the pseudo down spin sector, the Hamiltonian involves bond fermions only up to site $(N-1)$, leaving the $N$-th bond fermion completely free and thus at zero energy, which is the hallmark of the topological regime. In contrast, in the pseudo up spin sector the $N$-th bond fermion is coupled to the $(N-1)$-th bond fermion via the Zeeman field $V_z$, preventing the emergence of a decoupled zero energy mode. Further, Eq.~(\ref{eq:dnspin_majo}) allows us to identify the topological phase boundaries of the pseudo up and down spin chains. In particular, the pseudospin sector $H_\sigma$ enters the topological regime when the corresponding effective onsite energy satisfies the condition $\lambda_\pm \leq 2|\tilde{t}|$~\cite{Kitaev2001}. Thus, only the pseudo down sector hosts a decoupled zero energy bond fermion, the up sector remains gapped and parity nondegenerate for the parameters considered here. This behavior is reflected in the energy spectra of the two pseudospin sectors, shown in Figs.~\ref{fig:Energy_spin}(a) and \ref{fig:Energy_spin}(b). It is also important to emphasize that coupling of the $N$-th bond fermion does not, by itself, imply the absence of topology. In a finite system, it indicates edge hybridization that may shift the modes away from zero energy. In a topological phase, this splitting is expected to decrease with system size, vanishing in the thermodynamic limit. We illustrate this on an array of $N=3$ quantum dots, with $t=\Delta=1$ and $\alpha=1.2t$. Numerical results for the same set of parameters and larger quantum dot arrays, with up to \(N=100\) are presented later. When the pseudo down spin Hamiltonian is in the topological regime, the system exhibits a doubly degenerate ground state (GS), one in the odd parity subspace, $|\Psi_{GS}\rangle_\downarrow=|111\rangle_\downarrow$, in which all bond fermions are occupied, including zero energy one, and the other in the even parity subspace, $|\Psi_{GS}\rangle_\downarrow =|110\rangle_\downarrow$, which lacks the zero energy bond fermion. In addition, there are four excited states corresponding to the absence of one non zero energy bond fermion, with excitation energy $2|\tilde{t}|$, two in each parity sector. These spectra correctly reproduce the results for the Kitaev chain but in the bond fermion basis~\cite{Mohseni2023}. In contrast, the pseudo up spin sector has no parity degenerate ground state, a large even odd splitting, and the spectrum pushed to higher energy by $\lambda_+ >0$.

\begin{figure}
    \centering
    \includegraphics[width=\linewidth]{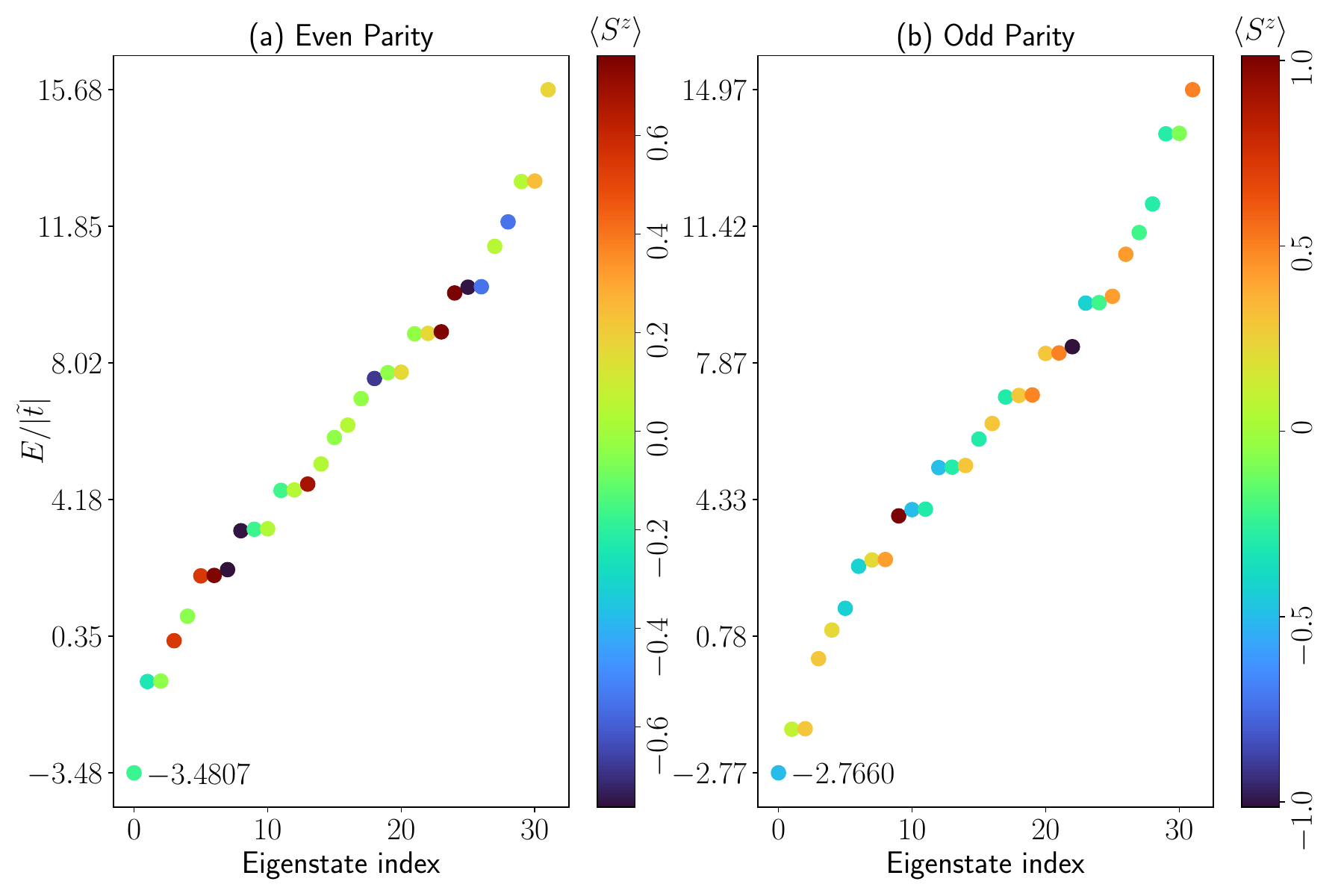}
    \caption{Energy spectra of the total Hamiltonian are calculated in the bond fermion basis. To enforce the pseudo down spin chain in the topological phase, we impose $\tilde{t} = \Delta_\alpha$ and $V_z = \sqrt{\mu^2 + \Delta^2}$. The parameters are $N=3$, $t=\Delta=1$ and $\alpha=1.2t$. The energies are normalized to $|\tilde{t}|$ and the color bars indicate the pseudospin polarizations of each eigenstate.}
    \label{fig:Energy_lambda_1}
\end{figure}

Let us now move to the two coupled effective Kitaev chain case, for this we are taking our full working Hamiltonian in the bond fermion basis as $H=H_\downarrow + H_\uparrow + H_{\downarrow \uparrow}$. Here we intentionally remove the term $-N(\mu + \tilde{\mu})$ as it only shifts the energies without affecting the physical interpretations of the energy spectra. Figure \ref{fig:Energy_lambda_1} shows the energy spectra in the (a) even and (b) odd fermion parity sectors for a finite chain of $N=3$ quantum dots in the presence of inter chain coupling. For these parameters, the even-odd parity splitting is $\sim 0.71|\tilde{t}|$, even though it is less than the first excitation energy $\sim 1.13|\tilde{t}|$. This indicates that the zero energy modes shift to finite energy for this choice of parameters. The color bars of Fig.~\ref{fig:Energy_lambda_1} represent the pseudospin polarization of the eigenstates. The low energy states are no longer fully spin polarized, instead exhibit a significant admixture of both pseudospin components. Although the ground states in the two parity sectors retain a partial down spin character, they are not eigenstates of a definite spin and therefore cannot be regarded as spin polarized modes. Consequently, the low energy excitations are not confined to a single pseudospin chain, but are instead delocalized across both pseudospin sectors. This underscores the importance of inter sector coupling in controlling the stability of Majorana induced parity degeneracy in realistic spinful $s$-wave proximity induced SM-SC array of QDs.

\begin{figure}
    \centering
    \includegraphics[width=\linewidth]{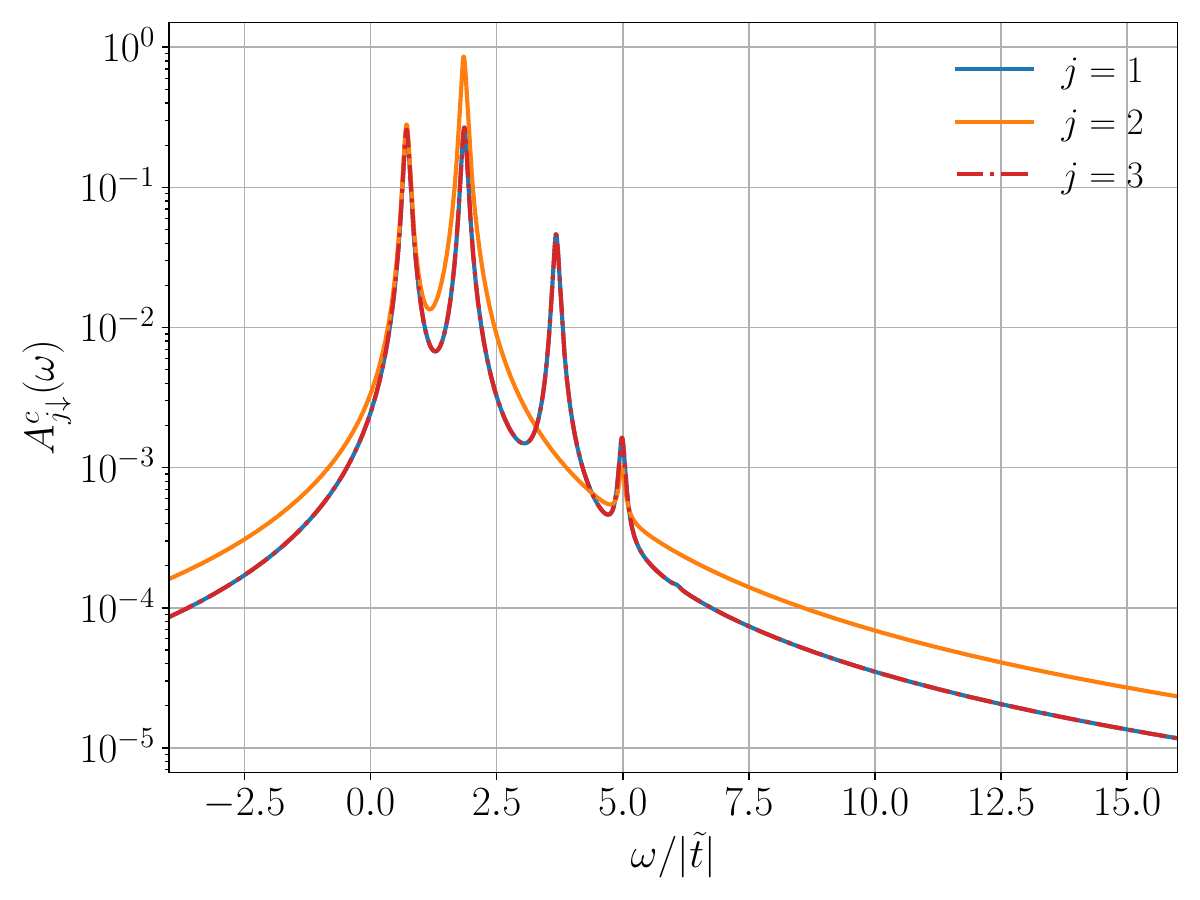}
    \caption{Site resolved real fermion spectral function $A_{j\downarrow}^c(\omega)$ for the down spin sector of a chain of three QDs, calculated using the electron creation operator $c_{j\downarrow}^\dagger$ on the bond fermion basis GS wave function, including a finite broadening $\eta$. The spectra are shown for the two boundary sites $j=1,3$ and the central site $j=2$. The parameters are same as Fig. \ref{fig:Energy_lambda_1}.}
    \label{fig:A_N3_real}
\end{figure}

Further, we compute the electron addition spectral function for the physical down spin fermions, distinguishing them from the pseudospin fermions. This quantity measures the spectral weight associated with adding a down spin fermion at site $j$ to the ground state. The local spectral function is defined as
\begin{equation}
    A_{j\downarrow}^c(\omega) = \sum_f |\langle \Psi_f|c_{j\downarrow}^\dagger|\Psi_{GS}\rangle|^2 \delta(\omega-(E_f - E_{GS})),
\end{equation}
where the sum runs over all the eigenstates $|\Psi_f\rangle$, including the ground state and all excited states, with corresponding energies $E_f$. The delta function enforces energy conservation during the particle addition process. In numerical calculations, we approximate it by a Lorentzian broadening, $\delta(\omega) = \frac{\eta}{\pi [\omega^2 + \eta^2]}$ with the broadening parameter $\eta = 0.05|\tilde{t}|$. Importantly, the spectral function is evaluated in the full Hilbert space, without restricting to a fixed fermion parity sector. The fermionic creation operator $c_{j\downarrow}^\dagger$ changes the total fermion number by one and therefore flips the fermion parity. Consequently, acting with $c_{j\downarrow}^\dagger$ on an even (odd) parity ground state produces a state in the odd (even) parity sector. Therefore, the spectral function decomposes into contributions from different parity sectors
\begin{align}
    \begin{split}
        A_{j\downarrow}^c(\omega) = &\sum_f |\langle \Psi_f^{odd}|c_{j\downarrow}^\dagger|\Psi_{GS}^{even}\rangle|^2 \delta(\omega-(E_f - E_{GS})) \\
        &+ \sum_f |\langle \Psi_f^{even}|c_{j\downarrow}^\dagger|\Psi_{GS}^{odd}\rangle|^2 \delta(\omega-(E_f - E_{GS})).
    \end{split}
\end{align}
Figure~\ref{fig:A_N3_real} presents the site resolved physical down spin electron addition spectral function $A_{j\downarrow}^{c}(\omega)$ for the three sites QD array. The resolved resonances correspond to transitions from the ground state to accessible excited states. Their positions are determined by the excitation energies, whereas their weights depend on the local transition matrix elements. The spectra therefore identify the excitations accessible through electron addition at each site and provide microscopic information relevant to tunneling spectroscopy. The site resolution is important, since Majorana zero modes are spatially localized near the ends of the chain~\cite{Thomale2013}. However, a zero energy peak at a boundary site alone establishes neither spatial localization nor topological character. A quantitative prediction of the differential conductance $dI/dV_{SD}$, where $V_{SD}$ denotes the source-drain voltage, additionally requires the coupling to the probe, its spin sensitivity, and the relevant transport processes.

The spectra at the boundary sites, $j=1$ and $j=3$, coincide, consistent with the equivalence of the two ends. The central site exhibits a different distribution of spectral weight. The lowest resolved resonance, at $\omega/|\tilde t|\approx0.71$, has substantial weight at all three sites. It therefore does not display a predominantly boundary localized down spin electron addition response. At higher energies, the spatial distribution varies between resonances. The peak near $\omega/|\tilde t|\approx1.8$ is enhanced at the central site. Stronger boundary enhanced peaks also appear at higher energies near $\omega/|\tilde t|\approx 3.7$ and $5$, respectively. These differences demonstrate that the spatial distribution of the spectral weight depends on the excitation being probed. Comparison with the up spin spectrum in Appendix~\ref{Appen:A} shows that both spin components contribute at low excitation energies. The lowest resonance is visible in both spin channels at the boundaries, while its central site contribution is much stronger for down spin electron addition. The resonances near $\omega/|\tilde t|\approx1.8$ and $3.7$ also exhibit contrasting site dependence in the two spin channels. Although the largest down spin peaks occur at lower energies than the largest up spin peak, the up spin response is not uniformly weaker. The distinction between the two spectra is therefore energy and site dependent redistribution of spectral weight, rather than a general suppression of one spin component. For the finite system shown here, the lowest resolved resonance occurs at a finite excitation energy and neither spectrum displays a resonance centered at $\omega=0$. These therefore establish a finite energy electron addition response with distinct spin and site dependence. Interpreting the lowest resonance as a zero energy Majorana mode requires additional evidence from its spatial structure and evolution with system size, the present $N=3$ spectra alone do not establish that identification.

\begin{figure}
    \centering
    \includegraphics[width=\linewidth]{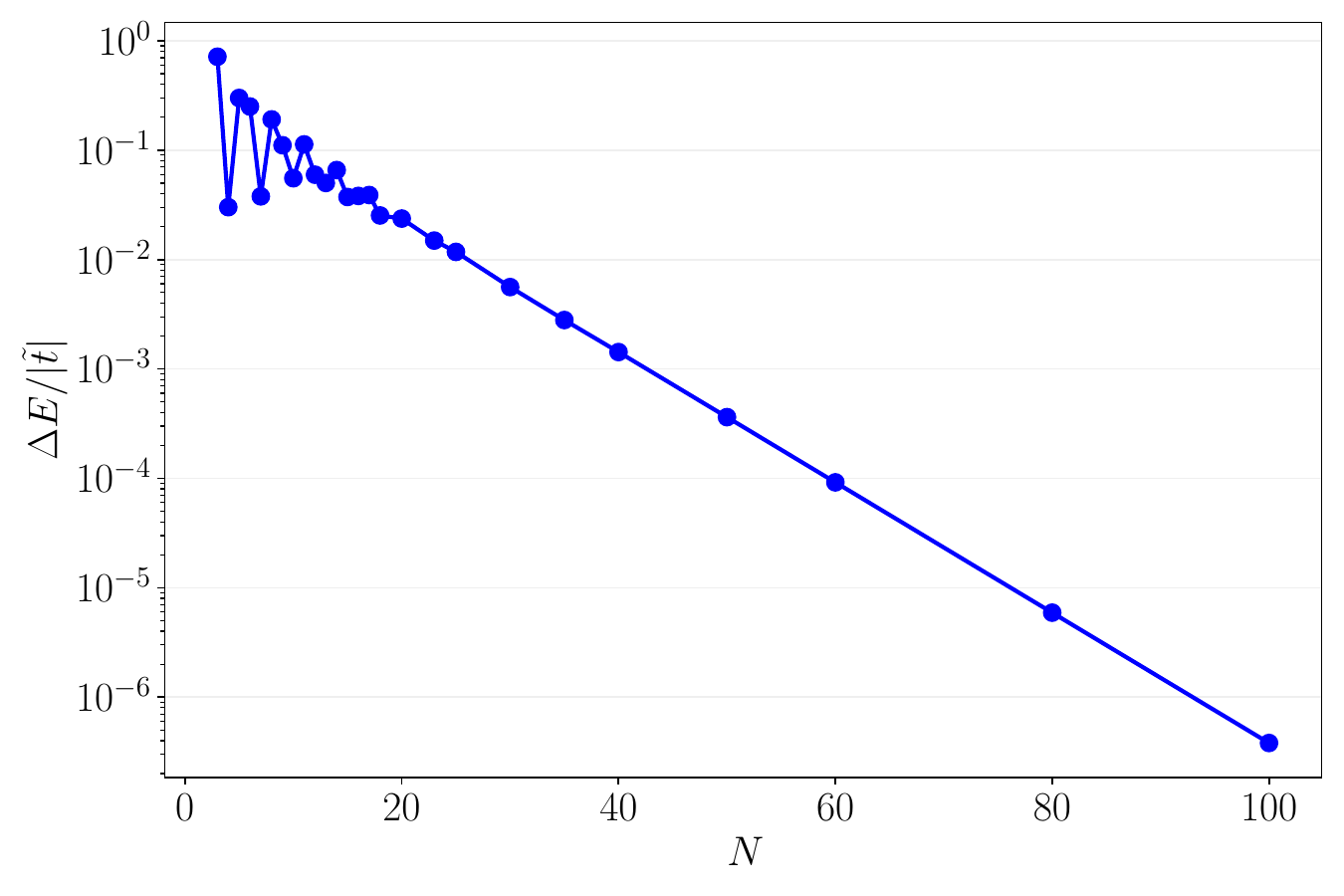}
    \caption{Finite size scaling of the even odd parity splitting $\Delta E=|E_0^{\rm{even}} - E_0^{\rm{odd}}|$ as a function of number of QDs in the chain obtain by diagonalizing the Bogoliubov--de Gennes (BdG) Hamiltonian.}
    \label{fig:DeltaE}
\end{figure}

Figure~\ref{fig:DeltaE} shows the system size dependence of the fermion parity splitting $\Delta E=|E_0^{\rm even}-E_0^{\rm odd}|$, where $E_0^{\rm{even}}$ and $E_0^{\rm{odd}}$ denote the ground state energies in the even and odd parity sectors, respectively for an open Kitaev chain describes in Eqs.~(\ref{eq:effective_bond}) for the Rashba SOC parameter $\alpha=1.2t$. We obtain $\Delta E$ by diagonalizing the Bogoliubov--de Gennes (BdG) Hamiltonian for chain lengths ranging from $N=3$ to $N=100$. For $N=3 \to 6$, we compare the lowest positive BdG eigenvalue ($\varepsilon_1$) with the parity splitting obtained by exact diagonalization of the many body Hamiltonian. For this quadratic Hamiltonian, occupying the lowest energy quasiparticle reverses the ground state parity at an energy cost of $\varepsilon_1$. The lowest positive BdG eigenvalue therefore gives the splitting $\Delta E =\varepsilon_1$. For short chains, $\Delta E/|\tilde t|$ varies nonmonotonically with $N$. For $N\gtrsim 20$, the data follow an approximately straight line on the semilogarithmic plot, consistent with an exponential decay $\Delta E\propto e^{-N/\xi}$. A fit over $20\leq N\leq 100$ gives a decay length of $\xi\simeq 7.3$ lattice sites. This size dependence is consistent with the decreasing overlap of Majorana modes localized at opposite ends of the chain and accounts for the appreciable parity splitting in short chains.

\section{Conclusions}
\label{Conclusions}
We studied a spinful one-dimensional semiconductor-superconductor model with Rashba spin orbit coupling, Zeeman splitting, and proximity induced $s$-wave pairing, motivated by recent realisation of quantum dot chains hybridized with superconductors. Using a local BdG transformation we obtained the Hamiltonian as describing two coupled Kitaev-like chains labeled by pseudospins. This allowed us to express the electronic properties  in terms of Majorana and bond fermions. With the exact solution for the Kitaev chain in bond fermions, the parity structure and low energy spectra were made transparent. Site resolved addition spectral functions exhibit MZM, whereas inter chain couplings lift the parity degeneracy and redistribute the spectral weight of MZM in finite systems. Future work  includes the effect of electron electron interactions and a microscopic, atomistic description of InAsP quantum dots in InP nanowires.

\begin{acknowledgments}
This work was supported by the Quantum Sensors Challenge Program QSP-078, High Throughput Networks HTSN-341, and Applied Quantum Computing AQC-004 Challenge Programs at the National Research Council of Canada, NSERC Discovery Grant No. RGPIN-2019-05714, NSERC PQS2D Alliance Quantum Grant No. ALLRP/578466-2022. This work was partly enabled by support provided by the Digital Research Alliance of Canada~\cite{alliancecan2025}. D. M. and P.H. acknowledge discussion with J. Danon, M. Leijnse. M. Korkusinski and A. W. Rodrigues.
\end{acknowledgments}

\bibliography{main}

@article{Kitaev2001,
  title={Unpaired majorana fermions in quantumwires},
  author={Kitaev, A Yu},
  journal={Physics-uspekhi},
  volume={44},
  number={10S},
  pages={131},
  year={2001},
  publisher={IOP Publishing},
doi     = {10.1070/1063-7869/44/10S/S29}
}

@article{Alicea2012,
  title={New directions in the pursuit of Majorana fermions in solid state systems},
  author={Alicea, Jason},
  journal={Reports on progress in physics},
  volume={75},
  number={7},
  pages={076501},
  year={2012},
  publisher={IOP Publishing},
doi     = {10.1088/0034-4885/75/7/076501}
}

@article{Beenakker2013,
  title={Search for Majorana fermions in superconductors},
  author={Beenakker, Carlo WJ},
  journal={Annu. Rev. Condens. Matter Phys.},
  volume={4},
  number={1},
  pages={113--136},
  year={2013},
  publisher={Annual Reviews},
doi     = {10.1146/annurev-conmatphys-030212-184337}
}

@article{Prada2020,
  author  = {Prada, E. and San-Jose, P. and de Moor, M. W. A. and Geresdi, A.
             and Lee, E. J. H. and Klinovaja, J. and Loss, D. and Nygard, J.
             and Aguado, R. and Kouwenhoven, L. P.},
  title   = {From Andreev to Majorana bound states in hybrid superconductor--semiconductor nanowires},
  journal = {Nat. Rev. Phys.},
  volume  = {2},
  pages   = {575},
  year    = {2020},
  doi     = {10.1038/s42254-020-0228-y}
}

@article{Nayak2008,
  title = {Non-Abelian anyons and topological quantum computation},
  author = {Nayak, Chetan and Simon, Steven H. and Stern, Ady and Freedman, Michael and Das Sarma, Sankar},
  journal = {Rev. Mod. Phys.},
  volume = {80},
  issue = {3},
  pages = {1083--1159},
  numpages = {0},
  year = {2008},
  month = {Sep},
  publisher = {American Physical Society},
  doi = {10.1103/RevModPhys.80.1083},
  url = {https://link.aps.org/doi/10.1103/RevModPhys.80.1083}
}

@article{Lutchyn2010,
  author  = {Lutchyn, R. M. and Sau, J. D. and Das Sarma, S.},
  title   = {Majorana fermions and a topological phase transition in semiconductor--superconductor heterostructures},
  journal = {Phys. Rev. Lett.},
  volume  = {105},
  pages   = {077001},
  year    = {2010},
  doi     = {10.1103/PhysRevLett.105.077001}
}

@article{Oreg2010,
  author  = {Oreg, Y. and Refael, G. and von Oppen, F.},
  title   = {Helical liquids and Majorana bound states in quantum wires},
  journal = {Phys. Rev. Lett.},
  volume  = {105},
  pages   = {177002},
  year    = {2010},
  doi     = {10.1103/PhysRevLett.105.177002}
}

@article{Stoudenmire2011,
  author  = {Stoudenmire, E. M. and Alicea, J. and Starykh, O. A. and Fisher, M. P. A.},
  title   = {Interaction effects in topological superconducting wires supporting Majorana fermions},
  journal = {Phys. Rev. B},
  volume  = {84},
  pages   = {014503},
  year    = {2011},
  doi     = {10.1103/PhysRevB.84.014503}
}

@article{Pan2020,
  title = {Physical mechanisms for zero-bias conductance peaks in Majorana nanowires},
  author = {Pan, Haining and Das Sarma, S.},
  journal = {Phys. Rev. Res.},
  volume = {2},
  issue = {1},
  pages = {013377},
  numpages = {33},
  year = {2020},
  month = {Mar},
  publisher = {American Physical Society},
  doi = {10.1103/PhysRevResearch.2.013377},
  url = {https://link.aps.org/doi/10.1103/PhysRevResearch.2.013377}
}

@article{DasSarma2021,
  author  = {Das Sarma, S. and Pan, H.},
  title   = {Disorder-induced zero-bias peaks in Majorana nanowires},
  journal = {Phys. Rev. B},
  volume  = {103},
  pages   = {195158},
  year    = {2021},
  doi     = {10.1103/PhysRevB.103.195158}
}

@article{Leijnse2012,
  author  = {Leijnse, M. and Flensberg, K.},
  title   = {Introduction to topological superconductivity and Majorana fermions},
  journal = {Semicond. Sci. Technol.},
  volume  = {27},
  pages   = {124003},
  year    = {2012},
  doi     = {10.1088/0268-1242/27/12/124003}
}

@article{Su2017,
  author  = {Su, Z. and Tacla, A. B. and Hocevar, M. and Car, D.
             and Plissard, S. R. and Bakkers, E. P. A. M.
             and Daley, A. J. and Pekker, D. and Frolov, S. M.},
  title   = {Andreev molecules in semiconductor nanowire double quantum dots},
  journal = {Nat. Commun.},
  volume  = {8},
  pages   = {585},
  year    = {2017},
  doi     = {10.1038/s41467-017-00665-7}
}

@article{Ptok2018,
  title   = {Quantum engineering of Majorana quasiparticles in one-dimensional optical lattices},
  author  = {Ptok, Andrzej and Cichy, Agnieszka and Doma{\'n}ski, Tadeusz},
  journal = {Journal of Physics: Condensed Matter},
  volume  = {30},
  pages   = {355602},
  year    = {2018},
  doi     = {10.1088/1361-648X/aad659}
}

@article{Dvir2022,
  author  = {Dvir, T. and Wang, G. and van Loo, N. and Liu, C.-X.
             and Mazur, G. P. and Bordin, A. and ten Haaf, S. L. D.
             and Wang, J.-Y. and van Driel, D. and Zatelli, F.
             and Li, X. and Malinowski, F. K. and Gazibegovic, S.
             and Badawy, G. and Bakkers, E. P. A. M. and Wimmer, M.
             and Kouwenhoven, L. P.},
  title   = {Realization of a minimal Kitaev chain in coupled quantum dots},
  journal = {Nature},
  volume  = {614},
  pages   = {445--450},
  year    = {2023},
  doi     = {10.1038/s41586-022-05585-1}
}

@Article{Haaf2024,
author={ten Haaf, Sebastiaan L. D.
and Wang, Qingzhen
and Bozkurt, A. Mert
and Liu, Chun-Xiao
and Kulesh, Ivan
and Kim, Philip
and Xiao, Di
and Thomas, Candice
and Manfra, Michael J.
and Dvir, Tom
and Wimmer, Michael
and Goswami, Srijit},
title={A two-site Kitaev chain in a two-dimensional electron gas},
journal={Nature},
year={2024},
month={Jun},
day={01},
volume={630},
number={8016},
pages={329-334},
issn={1476-4687},
doi={10.1038/s41586-024-07434-9},
url={https://doi.org/10.1038/s41586-024-07434-9}
}

@article{Bordin2025,
  author  = {Bordin, A. and Liu, C.-X. and Dvir, T. and Zatelli, F.
             and ten Haaf, S. L. D. and van Driel, D. and Wang, G.
             and van Loo, N. and Zhang, Y. and Wolff, J. C.
             and Van Caekenberghe, T. and Gazibegovic, S.
             and Bakkers, E. P. A. M. and Wimmer, M.
             and Kouwenhoven, L. P. and Mazur, G. P.},
  title   = {Enhanced Majorana stability in a three-site Kitaev chain},
  journal = {Nat. Nanotechnol.},
  volume  = {20},
  pages   = {726},
  year    = {2025},
  doi     = {10.1038/s41565-025-01894-4}
}

@article{Liu2024Gap,
  author  = {Liu, C.-X. and Bozkurt, A. M. and Zatelli, F.
             and ten Haaf, S. L. D. and Dvir, T. and Wimmer, M.},
  title   = {Enhancing the excitation gap of a quantum-dot-based Kitaev chain},
  journal = {Commun. Phys.},
  volume  = {7},
  pages   = {235},
  year    = {2024},
  doi     = {10.1038/s42005-024-01715-5}
}

@article{Samuelson2024,
  author  = {Samuelson, W. and Svensson, V. and Leijnse, M.},
  title   = {Minimal quantum dot based Kitaev chain with only local superconducting proximity effect},
  journal = {Phys. Rev. B},
  volume  = {109},
  pages   = {035415},
  year    = {2024},
  doi     = {10.1103/PhysRevB.109.035415}
}

@article{Haaf2025,
  author  = {ten Haaf, S. L. D. and Zhang, Y. and Wang, Q. and Bordin, A.
             and Liu, C.-X. and Kulesh, I. and Sietses, V. P. M.
             and Prosko, C. G. and Xiao, D. and Thomas, C.
             and Manfra, M. J. and Wimmer, M. and Goswami, S.},
  title   = {Observation of edge and bulk states in a three-site Kitaev chain},
  journal = {Nature},
  volume  = {641},
  pages   = {890},
  year    = {2025},
  doi     = {10.1038/s41586-025-08892-5}
}

@article{Mohseni2023,
  author  = {Mohseni, M. and Allami, H. and Miravet, D. and Gayowsky, D. J.
             and Korkusinski, M. and Hawrylak, P.},
  title   = {Majorana excitons in a Kitaev chain of semiconductor quantum dots in a nanowire},
  journal = {Nanomaterials},
  volume  = {13},
  pages   = {2293},
  year    = {2023},
  doi     = {10.3390/nano13162293}
}

@article{Pan2023,
  author  = {Pan, H. and Das Sarma, S.},
  title   = {Majorana nanowires, Kitaev chains, and spin models},
  journal = {Phys. Rev. B},
  volume  = {107},
  pages   = {035440},
  year    = {2023},
  doi     = {10.1103/PhysRevB.107.035440}
}

@article{Thomale2013,
  author  = {Thomale, R. and Rachel, S. and Schmitteckert, P.},
  title   = {Tunneling spectra simulation of interacting Majorana wires},
  journal = {Phys. Rev. B},
  volume  = {88},
  pages   = {161103},
  year    = {2013},
  doi     = {10.1103/PhysRevB.88.161103}
}

@article{Alicea2011,
  title   = {Non-Abelian statistics and topological quantum information processing in 1D wire networks},
  author  = {Alicea, J. and Oreg, Y. and Refael, G. and von Oppen, F. and Fisher, M. P. A.},
  journal = {Nature Physics},
  volume  = {7},
  pages   = {412--417},
  year    = {2011},
  doi     = {10.1038/nphys1915}
}

@article{BCS1957,
  title   = {Theory of superconductivity},
  author  = {Bardeen, J. and Cooper, L. N. and Schrieffer, J. R.},
  journal = {Physical Review},
  volume  = {108},
  pages   = {1175--1204},
  year    = {1957},
  doi     = {10.1103/PhysRev.108.1175}
}

@book{Tinkham,
  title     = {Introduction to Superconductivity},
  author    = {Tinkham, Michael},
  publisher = {McGraw-Hill},
  address   = {New York},
  year      = {1996},
  edition   = {2},
  isbn      = {9780486435039}
}

@book{DeGennes,
  title     = {Superconductivity of Metals and Alloys},
  author    = {de Gennes, P. G.},
  publisher = {Benjamin},
  address   = {New York},
  year      = {1966}
}

@article{Cygorek2020,
  author  = {Cygorek, M. and Korkusinski, M. and Hawrylak, P.},
  title   = {Atomistic theory of electronic and optical properties of InAsP/InP nanowire quantum dots},
  journal = {Phys. Rev. B},
  volume  = {101},
  pages   = {075307},
  year    = {2020},
  doi     = {10.1103/PhysRevB.101.075307}
}

@article{Manalo2021,
  author  = {Manalo, J. and Cygorek, M. and Altintas, A. and Hawrylak, P.},
  title   = {Electronic and magnetic properties of many-electron complexes in charged InAs$_x$P$_{1-x}$ quantum dots in InP nanowires},
  journal = {Phys. Rev. B},
  volume  = {104},
  pages   = {125402},
  year    = {2021},
  doi     = {10.1103/PhysRevB.104.125402}
}

@article{Dalacu2012,
  author  = {Dalacu, D. and Mnaymneh, K. and Lapointe, J. and Wu, X.
             and Poole, P. J. and Bulgarini, G. and Zwiller, V. and Reimer, M. E.},
  title   = {Ultraclean emission from InAsP quantum dots in defect-free wurtzite InP nanowires},
  journal = {Nano Lett.},
  volume  = {12},
  pages   = {5919},
  year    = {2012},
  doi     = {10.1021/nl303327h}
}

@article{Phoenix2022,
  author  = {Phoenix, J. and Korkusinski, M. and Dalacu, D. and Poole, P. J.
             and Zawadzki, P. and Studenikin, S. and Gaudreau, L.},
  title   = {Magnetic tuning of tunnel coupling between InAsP double quantum dots in InP nanowires},
  journal = {Sci. Rep.},
  volume  = {12},
  pages   = {5164},
  year    = {2022},
  doi     = {10.1038/s41598-022-08548-8}
}

@article{Manalo2024,
  title = {Microscopic design of a synthetic spin-1 chain in an InAsP quantum dot array},
  author = {Manalo, Jacob and Miravet, Daniel and Hawrylak, Pawel},
  journal = {Phys. Rev. B},
  volume = {109},
  issue = {8},
  pages = {085112},
  numpages = {8},
  year = {2024},
  month = {Feb},
  publisher = {American Physical Society},
  doi = {10.1103/PhysRevB.109.085112},
  url = {https://link.aps.org/doi/10.1103/PhysRevB.109.085112}
}

@article{Mohseni2025,
author = {Mohseni, Mahan and Cunha, Iann and Miravet, Daniel and Wania Rodrigues, Alina and Allami, Hassan and Assi, Ibsal and Korkusinski, Marek and Hawrylak, Pawel},
title = {Designing Majorana Quasiparticles in InAsP Quantum Dots in InP Nanowires with Variational Quantum Eigenvalue Solver},
journal = {physica status solidi (b)},
volume = {262},
number = {7},
pages = {2400552},
doi = {https://doi.org/10.1002/pssb.202400552},
url = {https://onlinelibrary.wiley.com/doi/abs/10.1002/pssb.202400552},
year = {2025}
}

@article{Benestad2024,
  author  = {Benestad, J. and Tsintzis, A. and Seoane Souto, R.
             and Leijnse, M. and van Nieuwenburg, E. and Danon, J.},
  title   = {Machine-learned tuning of artificial Kitaev chains from tunneling spectroscopy measurements},
  journal = {Phys. Rev. B},
  volume  = {110},
  pages   = {075402},
  year    = {2024},
  doi     = {10.1103/PhysRevB.110.075402}
}

@article{krawczyk2026,
  title = {AI-enhanced tuning of quantum dot Hamiltonians toward Majorana modes},
  author = {Krawczyk, Mateusz and Pawlowski, Jarolsaw},
  journal = {Phys. Rev. Appl.},
  volume = {25},
  issue = {6},
  pages = {064032},
  numpages = {13},
  year = {2026},
  month = {Jun},
  publisher = {American Physical Society},
  doi = {10.1103/xkbl-ctwn},
  url = {https://link.aps.org/doi/10.1103/xkbl-ctwn}
}

@article{Sau2012,
  title={Realizing a robust practical Majorana chain in a quantum-dot-superconductor linear array},
  author={Sau, Jay D and Sarma, S Das},
  journal={Nat. Commun.},
  volume={3},
  number={1},
  pages={964},
  year={2012},
  publisher={Nature Publishing Group UK London},
  doi = {10.1038/ncomms1966}
}

@article{Tsintzis2022Creating,
  title = {Creating and detecting poor man's Majorana bound states in interacting quantum dots},
  author = {Tsintzis, Athanasios and Souto, Rub\'en Seoane and Leijnse, Martin},
  journal = {Phys. Rev. B},
  volume = {106},
  issue = {20},
  pages = {L201404},
  numpages = {6},
  year = {2022},
  month = {Nov},
  publisher = {American Physical Society},
  doi = {10.1103/PhysRevB.106.L201404},
  url = {https://link.aps.org/doi/10.1103/PhysRevB.106.L201404}
}

@article{Souto2023,
  title = {Probing Majorana localization in minimal Kitaev chains through a quantum dot},
  author = {Souto, Rub\'en Seoane and Tsintzis, Athanasios and Leijnse, Martin and Danon, Jeroen},
  journal = {Phys. Rev. Res.},
  volume = {5},
  issue = {4},
  pages = {043182},
  numpages = {11},
  year = {2023},
  month = {Nov},
  publisher = {American Physical Society},
  doi = {10.1103/PhysRevResearch.5.043182},
  url = {https://link.aps.org/doi/10.1103/PhysRevResearch.5.043182}
}

@article{Zatelli2024,
  title={Robust poor man’s Majorana zero modes using Yu-Shiba-Rusinov states},
  author={Zatelli, Francesco and Van Driel, David and Xu, Di and Wang, Guanzhong and Liu, Chun-Xiao and Bordin, Alberto and Roovers, Bart and Mazur, Grzegorz P and Van Loo, Nick and Wolff, Jan C and others},
  journal={Nat. Commun.},
  volume={15},
  number={1},
  pages={7933},
  year={2024},
  doi     = {10.1038/s41467-024-52066-2}
}

@ARTICLE{Luethi2025,
       author = {{Luethi}, Melina and {Legg}, Henry F. and {Loss}, Daniel and {Klinovaja}, Jelena},
        title = "{Fate of poor man's Majoranas in the long Kitaev chain limit}",
      journal = {\prb},
         year = 2025,
        month = mar,
       volume = {111},
       number = {11},
          eid = {115419},
        pages = {115419},
          doi = {10.1103/PhysRevB.111.115419}
}

@article{Dourado2025,
  title = {Majorana sweet spots in three-site Kitaev chains},
  author = {Dourado, Rodrigo A. and Leijnse, Martin and Souto, Rub\'en Seoane},
  journal = {Phys. Rev. B},
  volume = {111},
  issue = {23},
  pages = {235409},
  numpages = {12},
  year = {2025},
  month = {Jun},
  publisher = {American Physical Society},
  doi = {10.1103/PhysRevB.111.235409},
  url = {https://link.aps.org/doi/10.1103/PhysRevB.111.235409}
}

@article{Bordin2026,
  title={Probing Majorana localization of a phase-controlled three-site Kitaev chain with an additional quantum dot},
  author={Bordin, Alberto and Bennebroek Evertsz’, Florian J and Roovers, Bart and Torres Luna, Juan D and Huisman, Wietze D and Zatelli, Francesco and Mazur, Grzegorz P and Ten Haaf, Sebastiaan LD and Badawy, Ghada and Bakkers, Erik PAM and others},
  journal={Nat. Commun.},
  volume={17},
  number={1},
  pages={2313},
  year={2026},
  publisher={Nature Publishing Group UK London},
  doi = {10.1038/s41467-026-68897-0}
}

@article{Yang2025,
  title = {Procedure to tune a three-site artificial Kitaev chain to host Majorana bound states based on transmon measurements},
  author = {Yang, Xiaozhou and Lyu, Zhaozheng and Wang, Xiang and Zhuo, Enna and Zhang, Yunxiao and Wang, Duolin and Shi, Yukun and Huang, Yuyang and Li, Bing and Song, Xiaohui and Li, Peiling and Tong, Bingbing and Dou, Ziwei and Shen, Jie and Liu, Guangtong and Qu, Fanming and Lu, Li},
  journal = {Phys. Rev. B},
  volume = {112},
  issue = {16},
  pages = {165418},
  numpages = {8},
  year = {2025},
  month = {Oct},
  publisher = {American Physical Society},
  doi = {10.1103/2r8x-9d9m},
  url = {https://link.aps.org/doi/10.1103/2r8x-9d9m}
}

@article{VanLoo2025,
  title={Single-shot parity readout of a minimal Kitaev chain},
  author={van Loo, Nick and Zatelli, Francesco and Steffensen, Gorm O and Roovers, Bart and Wang, Guanzhong and Van Caekenberghe, Thomas and Bordin, Alberto and van Driel, David and Zhang, Yining and Huisman, Wietze D and others},
  journal={Nature},
  volume={650},
  number={8101},
  pages={334--339},
  year={2026},
  publisher={Nature Publishing Group UK London},
  doi = {10.1038/s41586-025-09927-7}
}

@article{Koong2020,
  title = {Multiplexed Single Photons from Deterministically Positioned Nanowire Quantum Dots},
  author = {Koong, Zhe-Xian and Ballesteros-Garcia, Guillem and Proux, Rapha\"el and Dalacu, Dan and Poole, Philip J. and Gerardot, Brian D.},
  journal = {Phys. Rev. Appl.},
  volume = {14},
  issue = {3},
  pages = {034011},
  numpages = {11},
  year = {2020},
  month = {Sep},
  publisher = {American Physical Society},
  doi = {10.1103/PhysRevApplied.14.034011},
  url = {https://link.aps.org/doi/10.1103/PhysRevApplied.14.034011}
}

@article{jaworowski2017,
  title={Macroscopic singlet-triplet qubit in synthetic spin-one chain in semiconductor nanowires},
  author={Jaworowski, Blazej and Rogers, Nick and Grabowski, Marek and Hawrylak, Pawel},
  journal={Scientific Reports},
  volume={7},
  number={1},
  pages={5529},
  year={2017},
  publisher={Nature Publishing Group UK London},
  doi = {10.1038/s41598-017-05655-9}
}

@article{northeast2021,
  title={Optical fibre-based single photon source using InAsP quantum dot nanowires and gradient-index lens collection},
  author={Northeast, David B and Dalacu, Dan and Weber, John F and Phoenix, Jason and Lapointe, Jean and Aers, Geof C and Poole, Philip J and Williams, Robin L},
  journal={Scientific Reports},
  volume={11},
  number={1},
  pages={22878},
  year={2021},
  publisher={Nature Publishing Group UK London},
  doi = {10.1038/s41598-021-02287-y}
}

@article{Allami2025,
  title = {Two-qubit gate with macroscopic singlet-triplet qubits in synthetic spin-one chains in InAsP quantum dot nanowires},
  author = {Allami, Hassan and Miravet, Daniel and Korkusinski, Marek and Hawrylak, Pawel},
  journal = {Phys. Rev. B},
  volume = {111},
  issue = {11},
  pages = {115403},
  numpages = {13},
  year = {2025},
  month = {Mar},
  publisher = {American Physical Society},
  doi = {10.1103/PhysRevB.111.115403},
  url = {https://link.aps.org/doi/10.1103/PhysRevB.111.115403}
}

@unpublished{Islam2026prep,
  author = {Islam, Mijanur and Miravet, Daniel and Hawrylak, Pawel},
  title  = {Interaction dependent Majorana signatures and charge correlations in Rashba semiconductor-superconductor quantum dot arrays},
  note   = {in preparation},
  year   = {2026}
}

@article{Aasen2016,
  title = {Milestones Toward Majorana-Based Quantum Computing},
  author = {Aasen, David and Hell, Michael and Mishmash, Ryan V. and Higginbotham, Andrew and Danon, Jeroen and Leijnse, Martin and Jespersen, Thomas S. and Folk, Joshua A. and Marcus, Charles M. and Flensberg, Karsten and Alicea, Jason},
  journal = {Phys. Rev. X},
  volume = {6},
  issue = {3},
  pages = {031016},
  numpages = {28},
  year = {2016},
  month = {Aug},
  publisher = {American Physical Society},
  doi = {10.1103/PhysRevX.6.031016},
  url = {https://link.aps.org/doi/10.1103/PhysRevX.6.031016}
}

@article{Tsintzis2024,
  title = {Majorana Qubits and Non-Abelian Physics in Quantum Dot--Based Minimal Kitaev Chains},
  author = {Tsintzis, Athanasios and Souto, Rub{\'e}n Seoane and Flensberg, Karsten and Danon, Jeroen and Leijnse, Martin},
  journal = {PRX Quantum},
  volume = {5},
  issue = {1},
  pages = {010323},
  numpages = {25},
  year = {2024},
  month = {Feb},
  publisher = {American Physical Society},
  doi = {10.1103/PRXQuantum.5.010323},
  url = {https://link.aps.org/doi/10.1103/PRXQuantum.5.010323}
}

@article{CIORGA2001,
title = {Readout of a single electron spin based quantum bit by current detection},
journal = {Physica E: Low-dimensional Systems and Nanostructures},
volume = {11},
number = {1},
pages = {35-40},
year = {2001},
issn = {1386-9477},
doi = {https://doi.org/10.1016/S1386-9477(01)00153-9},
url = {https://www.sciencedirect.com/science/article/pii/S1386947701001539},
author = {M. Ciorga and A.S. Sachrajda and P. Hawrylak and C. Gould and P. Zawadzki and Y. Feng and Z. Wasilewski}
}

@article{Ciorga2002,
  title = {Collapse of the Spin-Singlet Phase in Quantum Dots},
  author = {Ciorga, M. and Wensauer, A. and Pioro-Ladriere, M. and Korkusinski, M. and Kyriakidis, J. and Sachrajda, A. S. and Hawrylak, P.},
  journal = {Phys. Rev. Lett.},
  volume = {88},
  issue = {25},
  pages = {256804},
  numpages = {4},
  year = {2002},
  month = {Jun},
  publisher = {American Physical Society},
  doi = {10.1103/PhysRevLett.88.256804},
  url = {https://link.aps.org/doi/10.1103/PhysRevLett.88.256804}
}

@article{Altintas2021,
  title = {Spin-valley qubits in gated quantum dots in a single layer of transition metal dichalcogenides},
  author = {Alt{\ifmmode \imath \else \i \fi{}}nta{\ifmmode \mbox{\c{s}}\else \c{s}\fi{}}, Abdulmenaf and Bieniek, Maciej and Dusko, Amintor and Korkusi{\ifmmode \acute{n}\else \'{n}\fi{}}ski, Marek and Pawlowski, Jaros{\l{}}aw and Hawrylak, Pawe{\l{}}},
  journal = {Phys. Rev. B},
  volume = {104},
  issue = {19},
  pages = {195412},
  numpages = {7},
  year = {2021},
  month = {Nov},
  publisher = {American Physical Society},
  doi = {10.1103/PhysRevB.104.195412},
  url = {https://link.aps.org/doi/10.1103/PhysRevB.104.195412}
}

@article{gaudreau2012,
  title={Coherent control of three-spin states in a triple quantum dot},
  author={Gaudreau, L and Granger, G and Kam, A and Aers, GC and Studenikin, SA and Zawadzki, P and Pioro-Ladriere, M and Wasilewski, ZR and Sachrajda, AS},
  journal={Nature Physics},
  volume={8},
  number={1},
  pages={54--58},
  year={2012},
  publisher={Nature Publishing Group UK London},
  doi = {10.1038/nphys2149},
  url = {https://doi.org/10.1038/nphys2149}
}

@article{Ciorga2000,
  title = {Addition spectrum of a lateral dot from Coulomb and spin-blockade spectroscopy},
  author = {Ciorga, M. and Sachrajda, A. S. and Hawrylak, P. and Gould, C. and Zawadzki, P. and Jullian, S. and Feng, Y. and Wasilewski, Z.},
  journal = {Phys. Rev. B},
  volume = {61},
  issue = {24},
  pages = {R16315(R)--R16318(R)},
  numpages = {0},
  year = {2000},
  month = {Jun},
  publisher = {American Physical Society},
  doi = {10.1103/PhysRevB.61.R16315},
  url = {https://link.aps.org/doi/10.1103/PhysRevB.61.R16315}
}

@article{Laferriere2021,
  title={Systematic study of the emission spectra of nanowire quantum dots},
  author={Laferriere, Patrick and Yeung, Edith and Korkusinski, Marek and Poole, Philip J and Williams, Robin L and Dalacu, Dan and Manalo, Jacob and Cygorek, Moritz and Altintas, Abdulmenaf and Hawrylak, Pawel},
  journal={Applied Physics Letters},
  volume={118},
  number={16},
  year={2021},
  publisher={AIP Publishing},
  doi = {10.1063/5.0045880},
  url = {https://doi.org/10.1063/5.0045880}
}

@misc{alliancecan2025,
  author       = {{Digital Research Alliance of Canada}},
  title        = {{Digital Research Alliance of Canada}},
  howpublished = {\url{https://alliancecan.ca}},
  year         = {2025},
  note         = {Accessed: 2025-07-26}
}


\appendix

\section{Spectral function of the up spin fermions}
\label{Appen:A}
\begin{figure}[H]
    \centering
    \includegraphics[width=\linewidth]{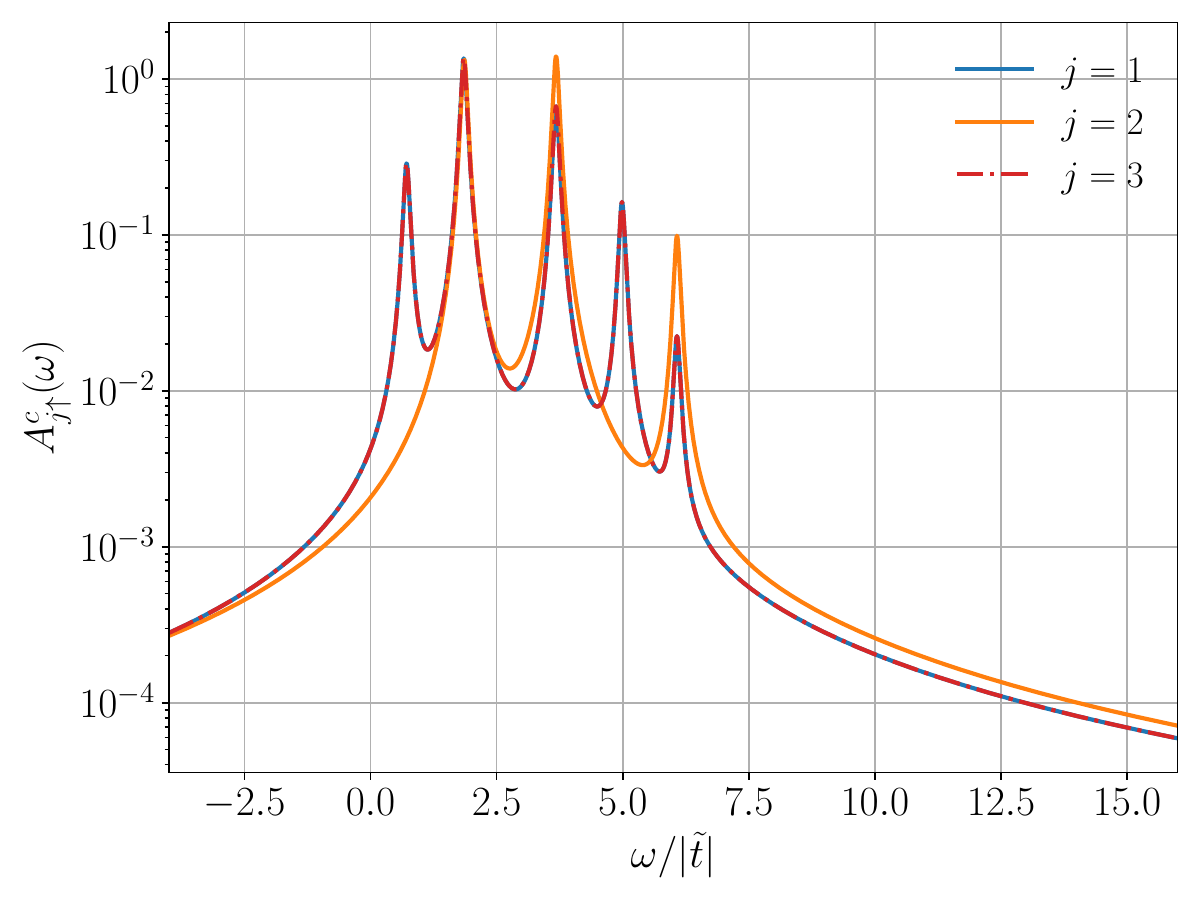}
    \caption{Site resolved spectral functions $A_{j\uparrow}^c(\omega)$ of the up spin fermions for the chain of $N=3$ quantum dots. The lowest energy spectral weight is primarily contributed by the edge fermions, while the contribution from the interior fermion appears at higher excitation energies.}
    \label{fig:A_up_N3}
\end{figure}

Figure~\ref{fig:A_up_N3} shows the site resolved up spin electron addition spectral function $A_{j\uparrow}^{c}(\omega)$ for the three site QD chain. As in the down spin spectrum, the responses at $j=1$ and $j=3$ coincide here also. The lowest resolved resonance, at $\omega/|\tilde t|\approx 0.71$, is pronounced at the boundaries and strongly suppressed at the central site. However, the central site already contributes a substantial peaks near $\omega/|\tilde t|\approx 1.85$ and $3.67$ along with the edge sites, while the resonance near $\omega/|\tilde t|\approx 5.0$ is enhanced at the boundaries. The strongest up spin resonances occur near $\omega/|\tilde t|\approx 1.85$ and $3.67$, where the later has its largest weight at $j=2$. Additional peak is visible near $\omega/|\tilde t|\approx 6.1$.

The up spin peak heights are comparable to, and for some resonances exceed, those in the down spin spectrum. Consequently, the comparison does not support a uniformly smaller up spin contribution. Instead, the two spin channels couple with different strengths to the accessible excitations at each site. The boundary enhancement of the lowest resonance in the up spin channel coexists with substantial central site weight in the down spin channel. Its spatial character therefore cannot be inferred from either spin resolved spectrum alone.

\end{document}